\documentclass[9pt,twocolumn,twoside,lineno]{pnas-new}

\templatetype{pnasmathematics} 

\title{Causal Scale Shift Associated with Phase Transition to Human Atrial Fibrillation}

\author[1,2,*]{Hiroshi Ashikaga}
\author[2]{Konstantinos N. Aronis} 
\author[2]{Susumu Tao}
\author[3]{Ryan G. James}
\affil[1]{IHU Liryc L'institut de rythmologie et mod\'{e}lisation cardiaque, H\^{o}pital Xavier Arnozan, Avenue du Haut L\'{e}v\^{e}que, 33604 Pessac cedex, France}
\affil[2]{Cardiac Arrhythmia Service, Johns Hopkins University School of Medicine, 600 N Wolfe Street, Carnegie 568, Baltimore, Maryland 21287, USA}
\affil[3]{Complexity Sciences Center, Department of Physics, University of California, Davis, One Shields Avenue, Davis, California 95616-8572, USA}

\leadauthor{Ashikaga} 

\authorcontributions{HA and RGJ conceived, designed and performed research, analyzed data, and wrote the manuscript. KNA and ST contributed to data analysis.}
\authordeclaration{No authors declare any conflict of interest.}
\correspondingauthor{\textsuperscript{1}To whom correspondence should be addressed. E-mail: hiroshi.ashikaga@ihu-liryc.fr}


\begin{abstract} 
An example of phase transition in natural complex systems is the qualitative and sudden change in the heart rhythm between sinus rhythm and atrial fibrillation (AF), the most common irregular heart rhythm in humans. While the system behavior is centrally controlled by the behavior of the sinoatrial node in sinus rhythm, the macro-scale collective behavior of the heart \textit{causes} the micro-scale behavior in AF. To quantitatively analyze this causation shift associated with phase transition in human heart, we evaluated the causal architecture of the human cardiac system using the time series of multi-lead intracardiac unipolar electrograms in a series of spatiotemporal scales by generating a stochastic renormalization group. We found that the phase transition between sinus rhythm and AF is associated with a significant shift of the peak causation from macroscopic to microscopic scales. Causal architecture analysis provides a quantitative tool to improve our understanding of causality in phase transitions in other natural and social complex systems.
\end{abstract}

\dates{This manuscript was compiled on \today}
\usepackage{wasysym}
\begin{document}

\verticaladjustment{-2pt}

\maketitle
\thispagestyle{firststyle}
\ifthenelse{\boolean{shortarticle}}{\ifthenelse{\boolean{singlecolumn}}{\abscontentformatted}{\abscontent}}{}

\dropcap{P}hase transitions between ordered and disordered states often reveal critical features of the underlying natual and social complex systems comprised of large numbers of components \citep{rvsole2011}. One example of phase transition in natural complex systems is the qualitative and sudden change in the heart rhythm between sinus rhythm and atrial fibrillation (AF) \citep{iravanian2017critical}, the most common irregular heart rhythm in human beings \citep{benjamin2018heart}. During sinus rhythm, the system behavior at the macroscopic scale is relatively simple, because it is centrally controlled by the behavior of the sinoatrial node. In contrast, as soon as the heart undergoes an order-disorder phase transition to AF, complex system behaviors emerge where the macro-scale collective behavior of the heart \emph{causes} the micro-scale behavior (\emph{`supersedence'}). The shift of causation from the centrally controlled sinus rhythm to AF, which controls the behaviors of individual cardiomyocytes to maintain itself, is clinically observable in the phenomenon called \textit{`AF begets AF'}, where a longer duration of pacing-maintained AF results in a longer maintenance of AF after cessation of pacing \citep{wijffels1995atrial}. Although this causation shift associated with phase transition to human AF has been qualitatively described, it has never been documented in a quantitative fashion.

The intra- and inter-scale interactions of multi-scale complex systems can be mathematically quantified by information theory \citep{allen2017multiscale}. In the cardiac system, we have previously shown that information-theoretic metrics such as mutual information \citep{shannon1948mathematical} and transfer entropy \citep{schreiber2000measuring} can quantify the interactions between micro-scale components \citep{ashikaga2015modelling}, and the interactions between micro- and macro-scale behaviors during fibrillation \citep{ashikaga2017inter}. We have also shown that those information-theoretic metrics can quantify how the interactions among micro-scale components alter the macro-scale collective behavior during fibrillation \citep{tao2017ablation} or cause an order-disoder phase transition to fibrillation \citep{ashikaga2018locating}. Recently, we have demonstrated that effective information, a quantity that captures causal interactions of a system between its unconstrained repertoire of possible cause and a specific state of possible effect \citep{tononi2003measuring}, can identify the causal scale of a numerical model of cardiac system during fibrillation \citep{ashikaga2018causal}.  

The aim of this study was to quantitatively analyze the causation shift associated with phase transition between sinus rhythm and AF in the human heart. To accomplish this aim, we used the paired time series data of unipolar electrograms in sinus rhythm and AF obtained by an intracardiac 64-lead basket catheter in patients \citep{tao2017ablation}, and described the cardiac system in a series of spatiotemporal scales by generating a renormalization group \citep{ashikaga2017inter,ashikaga2018causal}. Compared with numerical and experimental settings, invasive measurements in patients are limited by the relatively short data acquisition duration that is clinically feasible, and by the relatively small number of electrodes that can be inserted simultaneously in the beating human heart. To overcome those limitations, we developed stochastic renormalization, which allows statistical analysis of the aggregate of multiple random samples of the spatial and the temporal data from the entire time series and all the electrodes of the clinically acquired data in individual patients. Then we evaluated the causal architecture of the system in sinus rhythm and AF by quantifying effective information and causal capacity. Causal capacity is a system’s ability to transform interventions into effects in the maximal informative and efficacious manner \citep{hoel2017map}. We hypothesized that the phase transition between sinus rhythm and AF in human heart is associated with a significant shift in the scale of the peak causal power in the renormalization group.

\section*{Results}
\subsection*{Stochastic renormalization}
\begin{figure}[!h]
  \centering
  \includegraphics[width=0.8\linewidth,trim={1cm 11cm 4cm 1cm},clip]{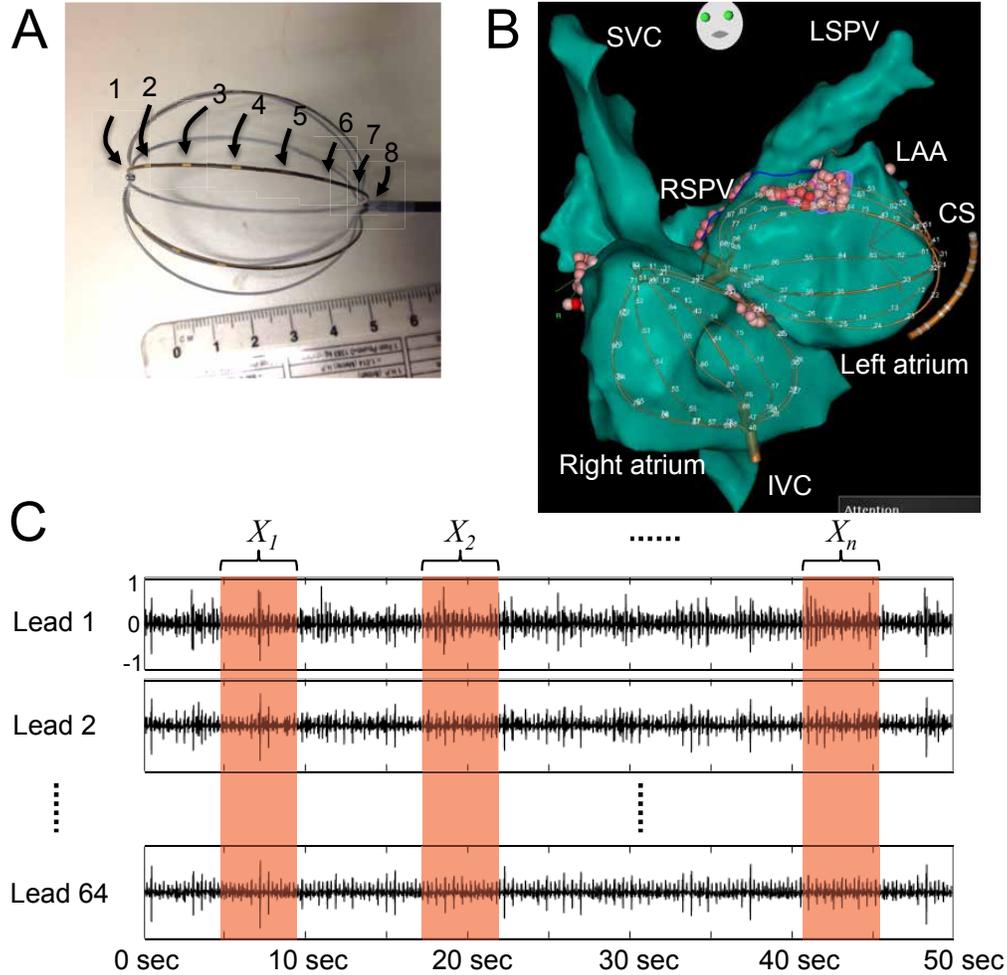}
  \caption{
    \textbf{Multi-lead mapping of the atria.} \textit{A. 64-lead basket catheter.} There are 8 leads (9-11 mm apart) per spline, and 8 splines (45 degree apart) per catheter, with a total of 64 (= 8 $\times$ 8) leads. \textit{B. Basket catheter configurations in the atria.} The basket catheter was inserted into the femoral vein and advanced to the right atrium through the inferior vena cava (IVC), and then was advanced trans-septally into the left atrium. In both panels, the electroanatomical shell of each atrium demonstrates that the catheter covers a wide field of view: LSPV, left superior pulmonary vein; RSPV, right superior pulmonary vein; SVC, superior vena cava; LAA, left atrial appendage; CS, coronary sinus catheter. \textit{C. Stochastic sampling of the time series.} The unipolar atrial electrograms during atrial fibrillation in leads 1, 2 and 64 of the basket catheter are shown. A total of $n$  time-series matrices of a fixed duration (= 4 seconds) were randomly sampled over the entire time series.  
  }
  \label{fig:basket}
\end{figure}

\begin{figure}[!h]
  \centering
  \includegraphics[width=0.8\linewidth,trim={1cm 4cm 1cm 0cm},clip]{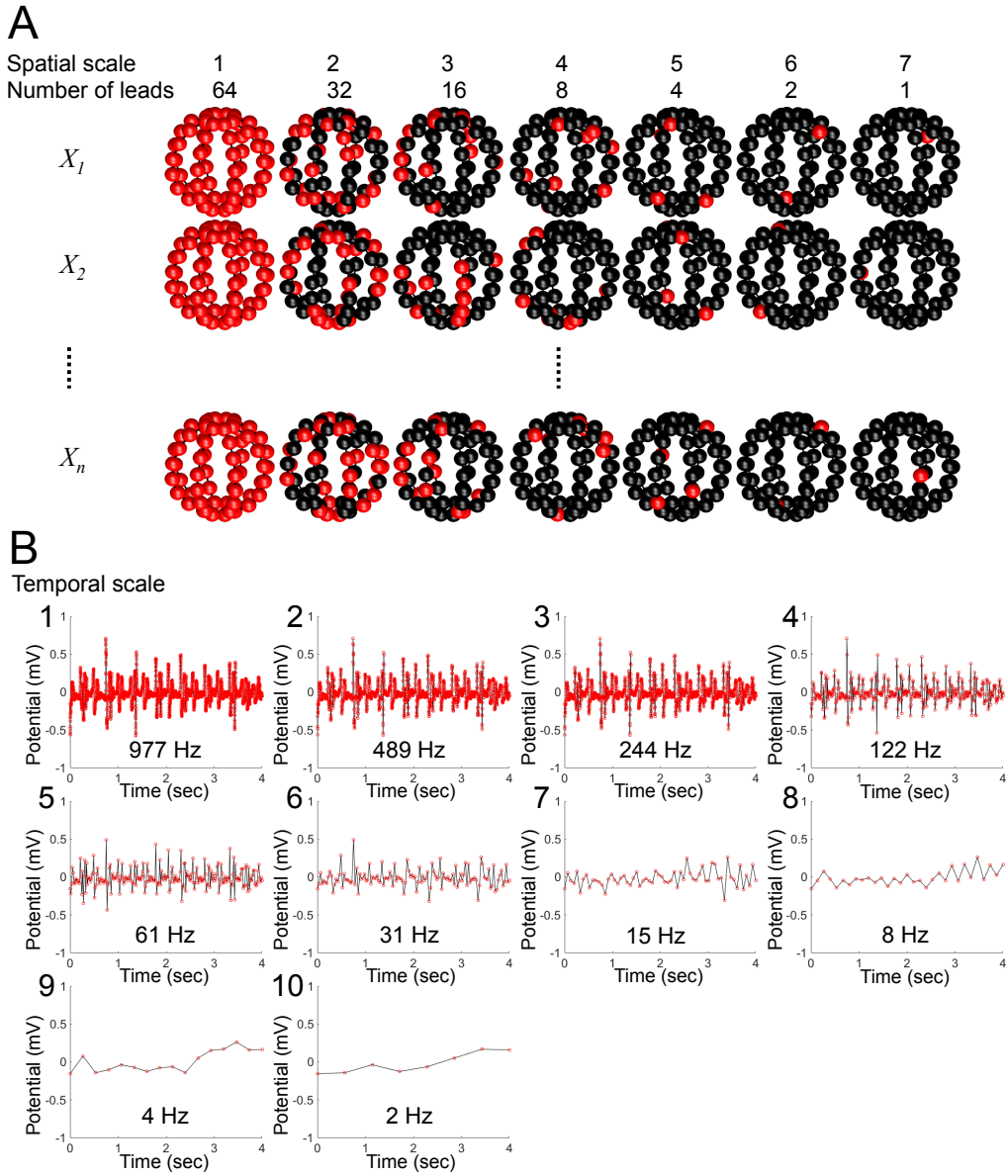}
  \caption{
    \textbf{Stochastic renormalization.} \textit{A. Spatial coarse-graining and rescaling.} For the time-series matrix $X_1$ (top row), we spatially coarse-grained and rescaled the original micro-scale description of the system (scale 1) by a factor of 2 by randomly sampling 32 (scale 2), 16 (scale 3), 8 (scale 4), 4 (scale 5), 2 (scale 6), and 1 (scale 7) lead(s). The leads shown in red indicate the sampled leads in each scale. We repeated the process by $n$ times. \textit{B. Temporal coarse-graining and rescaling.} For each time-series matrix $X_i$ of the randomly sampled lead in the original micro-scale description (scale 1), we temporally coarse-grained and rescaled the original micro-scale description of the system (scale 1; 977 Hz) by a factor of 2: scale 2 (488.5 Hz), scale 3 (244.25 Hz), scale 4 (122.125 Hz), scale 5 (61.0625 Hz), scale 6 (30.53125 Hz), scale 7 (15.265625 Hz), scale 8 (7.6328125 Hz), scale 9 (3.81640625 Hz), and scale 10 (1.908203125 Hz). The red circles indicate the sampled time points in each scale.
  }
  \label{fig:sp}
\end{figure}

First, to construct a series of robust and minimal macro-scale descriptions of the system, we generated a renormalization group of the system by a series of spatial and temporal transformation including coarse-graining and rescaling \cite{ashikaga2018causal}. Multi-lead mapping of the atria in sinus rhythm and AF was conducted by a 64-lead basket catheter (50 mm or 60 mm; Abbott Electrophysiology, Menlo Park, California; Figure~\ref{fig:basket}A), each atrium at a time (Figure~\ref{fig:basket}B).
Briefly, we extracted $n$ random samples of the spatial and the temporal data from the original micro-scale description of the system to 1) describe the causal architecture of the system over the entire time series; and 2) compensate for the relatively small number of leads (= 64) that can be inserted simultaneously in the beating human heart. First, we extracted a time series matrix $X_i$ of a fixed duration $\lambda$ from a uniformly distributed random time point in the time series (Figure~\ref{fig:basket}C). We chose $\lambda =$ 4 seconds, becasue the clinical standard of care uses a 4-second time window for diagnosis of AF configurations \cite{narayan2012treatment}. This process was repeated for $n$ times ($n=1, 10, 100, and 1000$) to generate a group of $n$ matrices ${X_1, X_2, ..., X_n}$. Second, for each time-series matrix $X_i$, we spatially coarse-grained the original micro-scale description of the system (scale 1) by a factor of 2 by randomly sampling 32 (scale 2), 16 (scale 3), 8 (scale 4), 4 (scale 5), 2 (scale 6), and 1 (scale 7) lead(s) (Fig.\ref{fig:sp}A). Third, for each time-series matrix $X_i$ of the randomly sampled lead in the original micro-scale description (scale 1), we temporally coarse-grained the original micro-scale description of the system (scale 1; 977 Hz) by a factor of 2: scale 2 (488.5 Hz), scale 3 (244.25 Hz), scale 4 (122.125 Hz), scale 5 (61.0625 Hz), scale 6 (30.53125 Hz), scale 7 (15.265625 Hz), scale 8 (7.6328125 Hz), scale 9 (3.81640625 Hz), and scale 10 (1.908203125 Hz) (Fig.\ref{fig:sp}B). 

\subsection*{Spatiotemporal representation}
\begin{figure}[!h]
  \centering
  \includegraphics[width=0.8\linewidth,trim={1cm 8cm 1cm 0cm},clip]{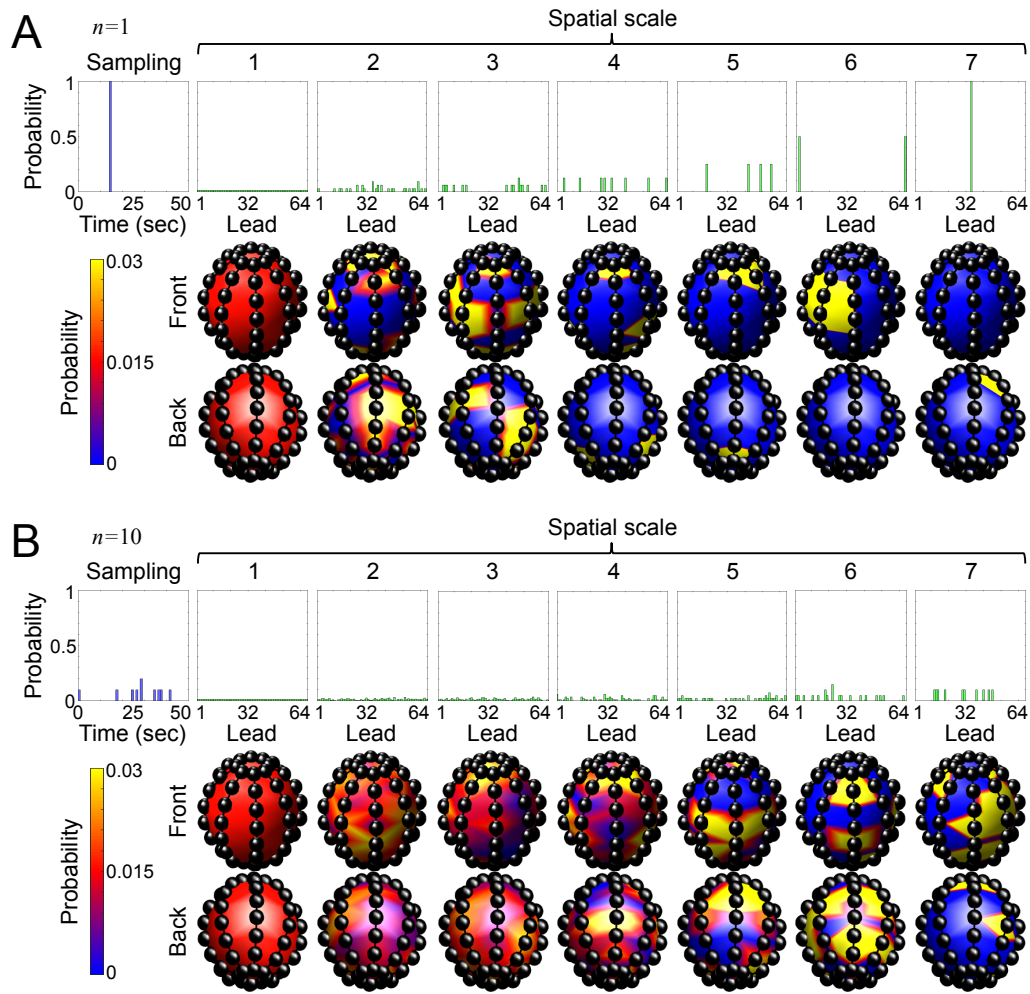}
  \caption{
    \textbf{Spatiotemporal representation.} \textit{A. $n=1$.} \textit{B. $n=10$.} The probability distribution of random time-series sampling (top left panel of respective subfigure) indicates the starting point of the time-series matrix $X_i$.
  }
  \label{fig:prob1}
\end{figure}

\begin{figure}[!h]
  \centering
  \includegraphics[width=0.8\linewidth,trim={1cm 8cm 1cm 0cm},clip]{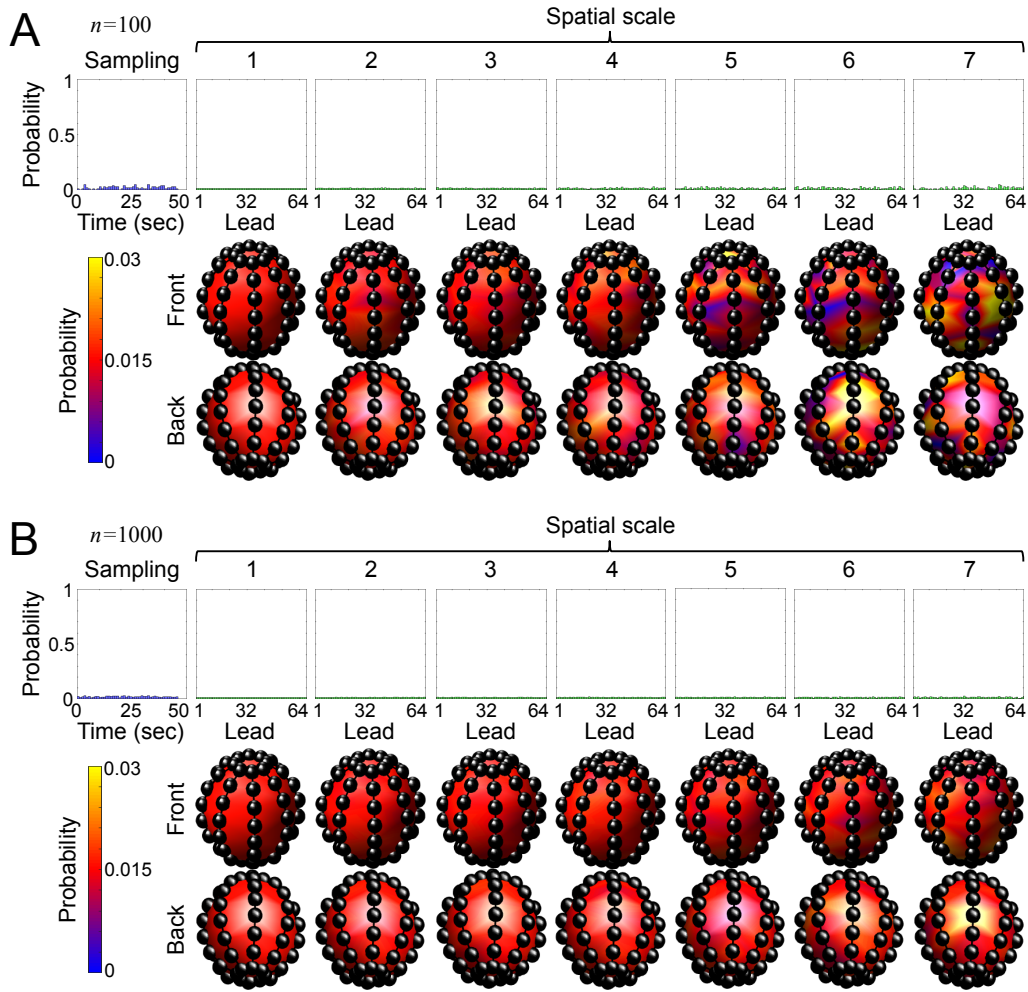}
  \caption{
    \textbf{Spatiotemporal representation. Continued.} \textit{A. $n=100$.} \textit{B. $n=1000$.} The probability distribution of random time-series sampling (top left panel of respective subfigure) indicates the starting point of the time-series matrix $X_i$.
  }
  \label{fig:prob2}
\end{figure}

Next, we assessed how much of the original micro-scale description of the system was represented spatially and temporally by each of the macro-scale descriptions of the system in the stochastic renormalization group. When $n=1$, there was only one time-series matrix with a probability of 1 (Fig.\ref{fig:prob1}A). In spatial scale 1, each lead had a uniform representation with a probability of 0.015625 (= 1/64). In spatial scale 2, approximately a half of the leads were represented but the other half were not. As the scale goes more macroscopic, fewer numbers of the leads were represented until only one lead was represented in scale 7. When $n=10$, ten time-series matrices were distributed over the entire time series (Fig.\ref{fig:prob1}B). Spatially, the leads showed a reasonably uniform representation from scale 1 to scale 3. Spatial representation beyond scale 4 showed missing leads. 
When $n=100$, a hundred time-series matrices were distributed over the entire time series (Fig.\ref{fig:prob2}A). Spatially, the leads showed a reasonably uniform representation from scale 1 to scale 5. Spatial representation beyond scale 6 showed missing leads. When $n=1000$, one thousand time-series matrices were distributed over the entire time series (Fig.\ref{fig:prob2}B). Spatially, the leads showed a reasonably uniform representation throughout all the scales. 

\subsection*{Change in the causal architecture between sinus rhythm and atrial fibrillation}
\begin{figure}[!h]
  \centering
  \includegraphics[width=0.8\linewidth,trim={1cm 7cm 3cm 0cm},clip]{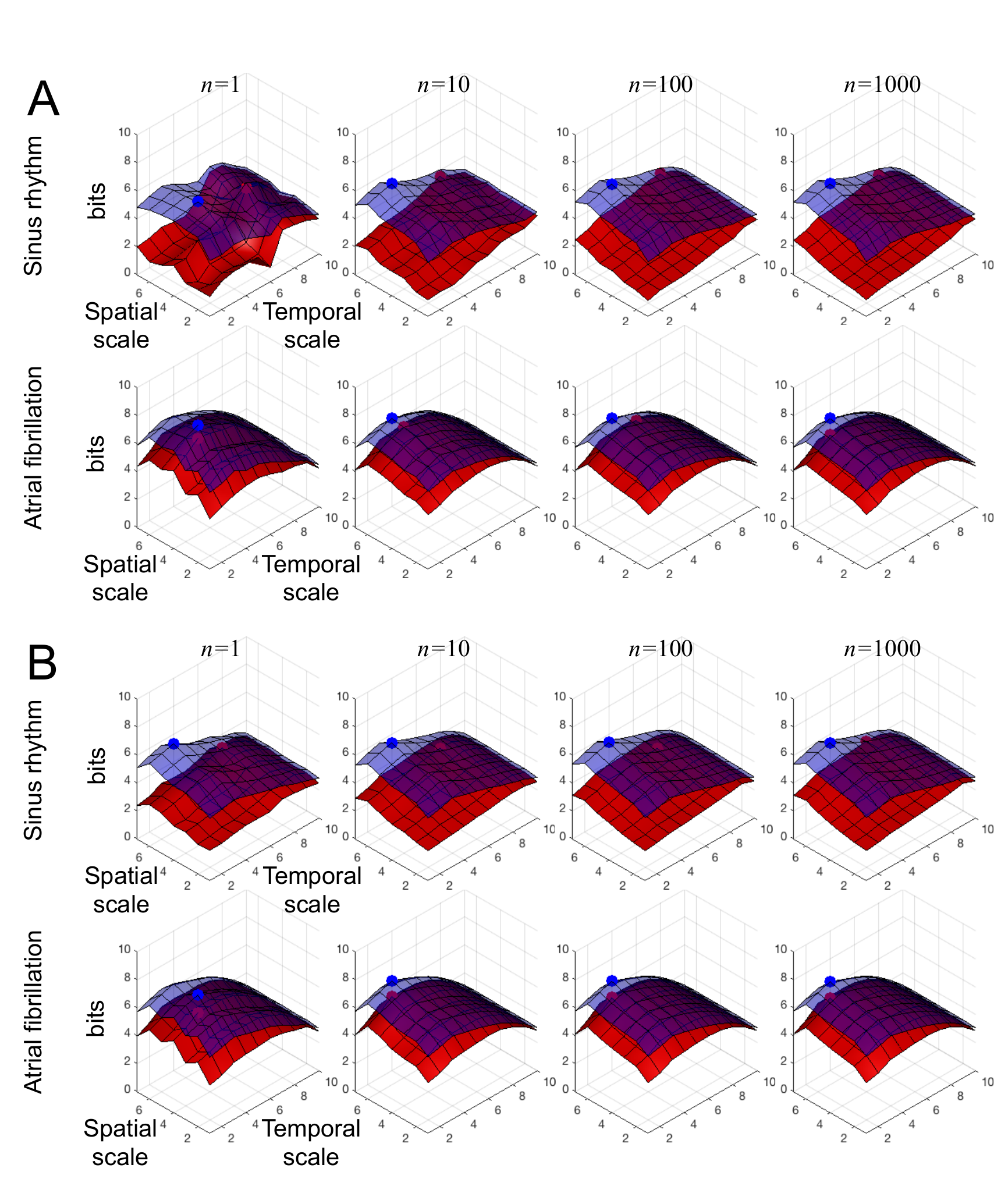}
  \caption{
    \textbf{Change in the causal architecture between sinus rhythm and atrial fibrillation.} In each panel, the red surface indicates the mean effective information (EI), and the blue surface the mean causal capacity (CC) over $n$ samples. The red circle indicates the scale of peak EI, and the blue circle the scale of peak CC. The unit is bits. \textit{A} shows a representative case, whereas in \textit{B}, the value of each component represents the mean EI and the mean CC of all patient cases. For both \textit{A} and \textit{B}, the top row indicates sinus rhythm, and the bottom atrial fibrillation. The columns indicate $n=1, 10, 100,$ and $1000$.
  }
  \label{fig:surface}
\end{figure}

\begin{figure}[!h]
  \centering
  \includegraphics[width=0.8\linewidth,trim={1cm 19cm 1cm 0cm},clip]{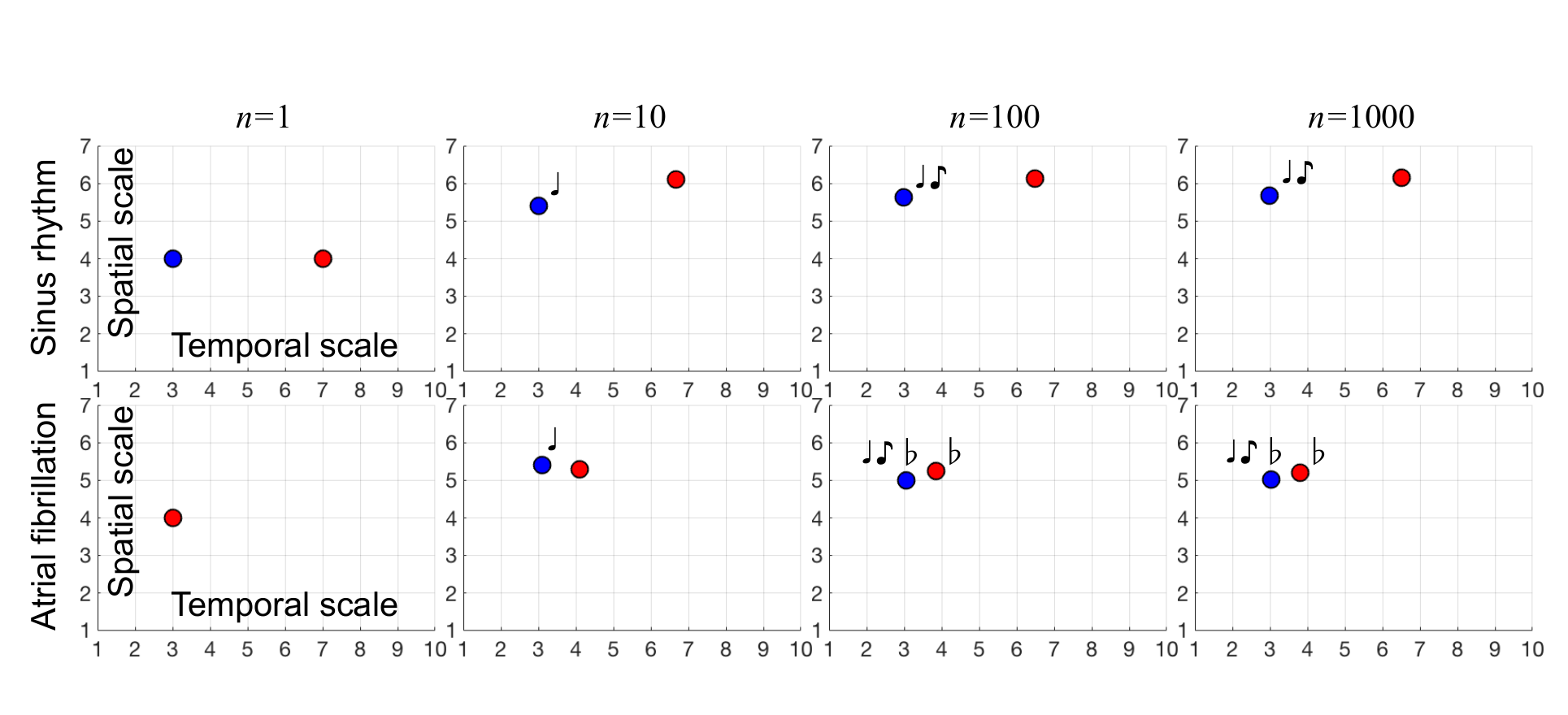}
  \caption{
    \textbf{Change in the peak causal scale between sinus rhythm and atrial fibrillation in a representative case.} The red circle indicates the mean scale of peak effective information (EI), and the blue circle the mean scale of peak causal capacity (CC). The top row indicates sinus rhythm, and the bottom atrial fibrillation. The columns indicate $n=1, 10, 100,$ and $1000$. \quarternote: p$<$0.05 \textit{vs.} EI (temporal scale); \eighthnote: p$<$0.05 \textit{vs.} EI (spatial scale); $\flat$: p$<$0.05 \textit{vs.} sinus rhythm (both temporal and spatial scales).
  }
  \label{fig:peaks}
\end{figure}

Then we evaluated the causal architecture of the system in sinus rhythm and AF. As expected, causal capacity was higher than effective information. As the number of samples $n$ increased (=1, 10, 100 and 1000), the surface of both causal capacity and effective information became smoother (Fig.\ref{fig:surface}A). In sinus rhythm, the scale of peak effective information was significantly more macroscopic than that of causal capacity in both the temporal and spatial scales at $n$=100 and 1000 (blue circles with symbol \quarternote and \eighthnote in Fig.\ref{fig:peaks}, top row). In AF, the scale of peak effective information was closer to, but still significantly more macroscopic than that of causal capacity in both the temporal and spatial scales at $n$=100 and 1000 (blue circles with symbol \quarternote and \eighthnote in Fig.\ref{fig:peaks}, bottom row). Between sinus rhythm and AF, the scale of peak causal capacity appeared similar (temporal scale ~ 3 and spatial scale ~ 6), but significantly different in both the temporal and spatial scales at $n$=100 and 1000 (blue circles with symbol $\flat$ in Fig.\ref{fig:peaks}). The scale of peak effective information in sinus rhythm was more macroscopic (temporal scale ~ 8 and spatial scale ~ 7) than that of AF (temporal scale ~ 3 and spatial scale ~ 6) (Fig.\ref{fig:surface}A), and the difference was statistically significant in both the temporal and spatial scales at $n$=100 and 1000 (red circles with symbol $\flat$ in Fig.\ref{fig:peaks}). This finding indicates a shift of causal scale from simple system behaviors with central control (sinus rhythm) to complex system behaviors with no central control (AF). The mean of the pooled data from all the seven patient cases showed essentially the same result (Fig.\ref{fig:surface}B).

\subsection*{Change in the system utilization between sinus rhythm and atrial fibrillation}
\begin{figure}[!h]
  \centering
  \includegraphics[width=0.8\linewidth,trim={1cm 15cm 1cm 0cm},clip]{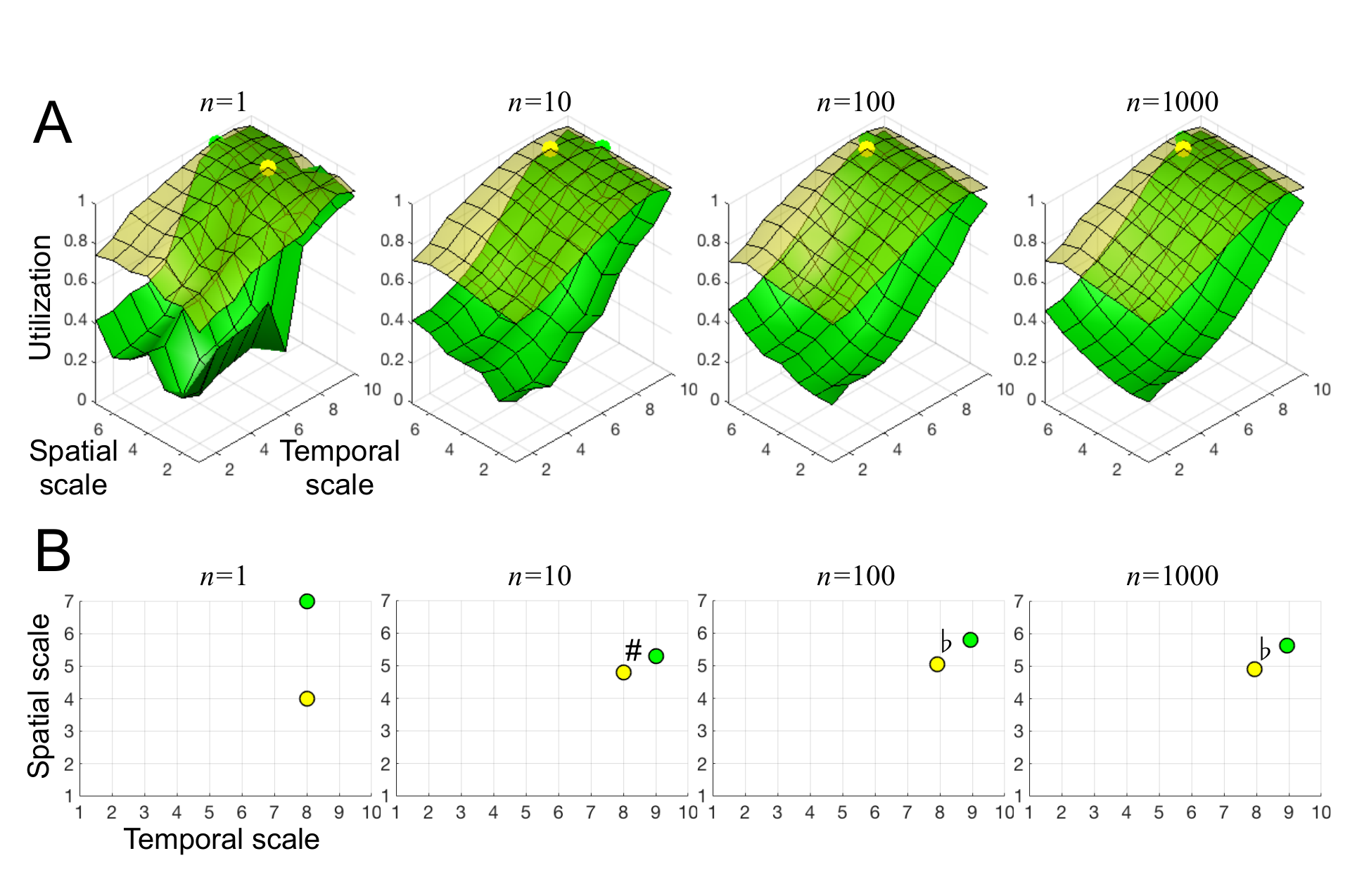}
  \caption{
    \textbf{Change in the system utilization between sinus rhythm and atrial fibrillation in a representative case.} \textit{A. System utilization.} The green surface indicates the mean system utilization in sinus rhythm, and the yellow surface the mean system utilization in atrial fibrillation over $n$ samples. The green and the yellow circles indicate the scales of peak system utilization in sinus rhythm and atrial firbillaion, respectively. \textit{B. Peak scale.} The green circle indicates the mean scale of peak system utilization in sinus rhythm, and the yellow circle the mean scale of peak system utilization in atrial fibrillation. For both panels A and B, the columns indicate $n=1, 10, 100,$ and $1000$. $\sharp$: p$<$0.05 \textit{vs.} sinus rhythm (temporal scale); $\flat$: p$<$0.05 \textit{vs.} sinus rhythm (both temporal and spatial scales).
  }
  \label{fig:util}
\end{figure}

Lastly, we compared how much of the system's ability was actually utilized in sinus rhythm and AF. Overall, as the number of samples $n$ increased (=1, 10, 100 and 1000), the surface of system utilization in both sinus rhythm and AF became smoother (Fig.\ref{fig:util}A). In sinus rhythm, the system utilization was higher in macroscopic scales with the peak in temporal scale ~ 9 and spatial scale ~ 6. In AF, the system utilization was also higher in macroscopic scales with the peak in temporal scale ~ 8 and spatial scale ~ 5. Between sinus rhythm and AF, the scale of peak system utilization was close togther, but significantly different in both the temporal and spatial scales at $n$=100 and 1000 (yellow circles with symbol $\flat$ in Fig.\ref{fig:util}B).

\section*{Discussion}
\subsection*{Main findings}
First, we found that stochastic renormalization provided a systematic means of extracting macro-scale features and reducing the number of degrees of freedom while overcoming the limitations of clinically available data by random sampling of multiple spatial and temporal segments of the system. The sensitivity analysis indicates that the spatial and temporal representation of the system was close to homogeneous at $n$=1000 (Fig.\ref{fig:prob2}B).

Second, the phase transition between sinus rhythm and AF in human heart was associated with a significant shift of the peak causation, quantified by effective information, from macroscopic to microscopic scales. The scale of peak effective information in sinus rhythm was more macroscopic (temporal scale ~ 8 and spatial scale ~ 7) than that of AF (temporal scale ~ 3 and spatial scale ~ 6) (Fig.\ref{fig:surface}A), and the difference was statistically significant in both the temporal and spatial scales at large samples (red circles with symbol $\flat$ in Fig.\ref{fig:peaks}).

Third, the phase transition between sinus rhythm and AF in human heart was also associated with a significant shift in the scale of peak causal capacity, but the magnitude of shift was smaller than that of effective information (blue circles with symbol $\flat$ in Fig.\ref{fig:peaks}). This suggests that causal capacity captures the same underlying governing dynamics of the two different phases, sinus rhythm and AF, in the cardiac system.

Lastly, system utilization was significantly higher in AF than in sinus rhythm at all spatiotemporal scales (Fig.\ref{fig:util}A). This finding indicates that the phase transition from sinus rhythm to AF in human heart is accompanied by a rise in the information content generated with reference to its systam capacity. It also showed that the scale of peak system utilization was significantly more macroscopic in sinus rhythm than in AF both spatially and temporally (yellow circles with symbol $\flat$ in Fig.\ref{fig:util}B). This finding is consistent with the causal scale shift from centrally controlled behaviors to complex system behaviors.

\subsection*{Clinical implications}
In this study, we provide quantitative evidence of causation shift associated with phase transition to human AF. Our findings have two important clinical implications. First, our analysis sheds new light on the mechanism that maintains AF, which could be organized reentrant circuits (`rotors')  \citep{mandapati2000stable,narayan2012treatment,haissaguerre2014driver}, multiple wavelets \citep{moe1959atrial,allessie2010electropathological,de2016direct} or focal activities \citep{lee2013high}, but remains elusive. We found that the human cardiac system in AF has a spatiotemporal scale at which effective information peaks (Fig.\ref{fig:surface}). The most causal power of the system does not lie in the most microscopic nor the most macroscopic scale, which allows coexistence of both downward and upward causation in the system. This finding is precisely the same as that of numerical model of cardiac system with rotors \citep{ashikaga2018causal}. Therefore, our result is consistent with the hypothesis that AF is maintained by rotors. However, our result does not rule out the possibilty that AF can also be maintained by multiple wavelets or focal activities. Second, our analysis provides a new approach to quantifying AF. We quantified the causal architecture of human AF, rather than simply the presence or absence of AF. Quantitative analysis of the causal scale of human AF may provide patient-specific diagnostic parameters that could potentially serve as a valid endpoint for therapeutic interventions.

\subsection*{Implications for other complex systems}
Phase transitions between ordered and disordered states are often observed in complex systems that consist of large numbers of interacting components \citep{buhl2006disorder,miramontes1995order}. Examples of phase transitions include network synchronization \citep{nishikawa2010network}, climate changes \cite{lenton2008tipping}, stock market crashes \cite{may2008complex}, and flocking of birds \cite{Christodoulidi2014flocking}. Our results demonstrate that the phase transition between ordered (sinus rhythm) and disordered (AF) states in the human heart is associated with a significant causation shift in the system behavior. This finding indicates a possibility that phase transitions in other natural and social complex systems may also be associated with a causation shift. In other words, the causal scale could be used as an order parameter to define phase transitions in complex systems. Our study demonstrates the value of the heart as a generic model to improve our understanding of causal architectures associated with phase tranisitions in natural and social complex systems.

\subsection*{Limitations}
This was a single-center, retrospective study with a relatively small sample size. Therefore, there is a non-negligible chance of selection bias. Our results should be confirmed in a larger prospective study. In addition, our analysis might have been affected by the location of the basket catheter within the heart chamber \citep{benharash2015quantitative}. However, our results showed consistent causal architectures in sinus rhythm and AF in both atria even though the measurements in different heart chambers were acquired at different time points. Therefore, we believe that the causal architecture analysis is spatially and temporally robust. 

\subsection*{Conclusions}
The phase transition between sinus rhythm and AF in human is associated with a significant shift in the causal scale. Effective information and causal capacity analysis provides a quantitative tool to improve our understanding of a change in causal architectures associated with phase transitions in natural and social complex systems.

\matmethods{
We performed the data analysis using Matlab R2018a (Mathworks, Inc.).

\subsection*{Multi-lead mapping of atrial fibrillation in patients}
We analyzed 7 paired time series data sets in sinus rhythm and AF out of 6 patients who were referred for the first catheter ablation of persistent AF. None of the patients had had prior cardiac procedures. A 64-lead basket catheter (50 or 60 mm; Abbott Electrophysiology, Menlo Park, California, USA) was advanced to the right atrium via the inferior vena cava or the left atrium trans-septally. Unipolar electrograms from the basket catheter were filtered at 0.05 to 500 Hz and recorded during AF for 60 seconds at a sampling frequency of 977 Hz (Cardiolab; GE Healthcare, Waukesha, WI). The far-field QRS-T complexes were removed, and the time series was divided into 5 consecutive 10-second time windows, as described previously \citep{tao2017ablation}. 

\subsection*{Effective information}
We treated each lead on the basket catheter as a time-series process $X$. \textit{Entropy} $H$ of each time-series process $X$ is
\begin{equation}
H(X)=-\sum_{x}p(x)\log_{2}p(x)
\label{eq:entropy}
\end{equation}
where $p(x)$ denotes the probability mass function of the time series generated by $X$. \textit{Effective information} quantifies the information generated when the system enters a specific state of possible effect $Y$ out of its unconstrained probability distribution of possible cause $X$ \citep{tononi2003measuring}.
\begin{eqnarray}
EI(X\rightarrow Y)&=&I(X;Y)\\
&=&H(X)+H(Y)-H(X,Y)\\
&=&\sum_{x,y}p(x,y)\log_{2}\frac{p(x,y)}{p(x)p(y)}
\label{eq:mi}
\end{eqnarray}
where $X$ has a uniform probability distribution so that it provides the maximum entropy $H(X)_{max}$ \citep{jaynes1957information}. $I(X;Y)$ is mutual information, $p(x,y)$ and $H(X,Y)$ denote the joint probability mass function and the joint entropy of $X$ and $Y$, respectively. Derivation of effective information in cardiac systems has been described elsewhere \citep{ashikaga2018causal}. Briefly, we binned the unipolar electrograms of $n$ leads at each scale ($n=1, 2, 4, 8, 16, 32, 64$) into a histogram of $a$ bins according to the potential amplitude. At each time point, the unconstrained repertoire of all possible causes $X$ consists of $a^n$ possible states with equal probability ${1/{a^n}}$. Then the entire time series $X$ can be defined as a vector of uniformly distributed numbers between 0 and $a^n-1$. Similarly, the time series $Y$ can be defined as a vector of decimal numbers between 0 and $a^n-1$, each of which represents a specific state of possible effect derived from the patient data. To calculate the probability distribution of $X$ and $Y$, we binned $X$ and $Y$ into a histogram of $b$ bins ($b<a^n$) according to the decimal numbers between 0 and $a^n-1$. To minimize the computational cost, we used $a=2^{7}=128$ and $b=2^{10}=1,024$ in this study. 

\subsection*{Causal capacity}
\textit{Causal capacity} quantifies a system’s ability to transform the cause $X$ into the effect $Y$ in the maximal informative and efficacious manner \citep{hoel2017map}. Causal capacity is equal to channel capacity between the cause $X$ and the effect $Y$ of the system, which is the maximum of mutual information where the maximization is with respect to the distribution of the cause $X$ \citep{2006:CoverEIT}.
\begin{eqnarray}
CC(X\rightarrow Y)&=&\max_{p(x)}EI(X\rightarrow Y)\\
&=&\max_{p(x)}I(X;Y)\\
\label{eq:cc}
\end{eqnarray}
We used the algorithm developed by Blahut \citep{blahut1972computation} and Arimoto \citep{arimoto1972algorithm} to compute channel capacity.
\subsection*{System utilization}
\textit{System utilization} quantifies how much of the system's ability is actually utilized when transforming the cause $X$ into the effect $Y$.
\begin{equation}
u(X\rightarrow Y)=\frac{EI(X\rightarrow Y)}{CC(X\rightarrow Y)}
\label{eq:entropy}
\end{equation}
System utilization is a novel extension to integrated information theory \citep{tononi2016integrated}.
}

\showmatmethods 

\acknow{This work was supported by the Fondation Leducq Transatlantic Network of Excellence (to H.A.).}

\showacknow 

\pnasbreak

\bibliography{fib_ref}

\end{document}